\documentclass[preprint,showpacs,preprintnumbers,floatfix]{revtex4}
\usepackage{amsmath}
\usepackage{amssymb}
\usepackage{graphicx}
\usepackage{dcolumn}
\usepackage{bm}
\usepackage[dvips]{color}

\begin{document}

\title{ Evidence for Antiferromagnetic Order in La$_{2-x}$Ce$_{x}$CuO$_{4}$
from Angular Magnetoresistance Measurements }
\author{K. Jin, X. H. Zhang, P. Bach, and R. L. Greene}
\affiliation{Center for Nanophysics $\And $ Advanced Materials and Department of Physics,
University of Maryland, College Park, MD 20742, USA}
\date{\today }

\begin{abstract}
We investigated the in-plane angular magnetoresistivity (AMR) of $%
T^{^{\prime }}$-phase La$_{2-x}$Ce$_{x}$CuO$_{4}$ (LCCO) thin films ($%
x=0.06-0.15$) fabricated by a pulsed laser deposition technique. The
in-plane AMR with $\mathbf{H}\parallel ab$ shows a twofold symmetry instead
of the fourfold behavior found in other electron-doped cuprates such as Pr$%
_{2-x}$Ce$_{x}$CuO$_{4}$ and Nd$_{2-x}$Ce$_{x}$CuO$_{4}$. The twofold AMR
disappears above a certain temperature, $T_{D}$. The $T_{D}(x)$ is well
above $T_{c}(x)$ for $x=0.06$ ($\sim 110$ K), and decreases with increasing
doping, until it is no longer observed above $T_{c}(x)$ at $x=0.15$. This
twofold AMR below $T_{D}(x)$ is suggested to originate from an
antiferromagnetic or spin density wave order.
\end{abstract}

\maketitle

High$-T_{c}$ superconductivity in the cuprates can be induced in the parent
antiferromagnetic (AFM) insulator by the doping of either holes or
electrons, corresponding to the formation of so-called hole- or
electron-doped high$-T_{c}$ cuprates. Neutron-scattering experiments have
revealed that the N\'{e}el ordering is rapidly suppressed in the hole-doped
cuprates \cite{RMP-70-897} but persists to much higher doping levels in the
electron-doped systems \cite{Nat-423-522,Nat-445-186}. For the frequently
studied hole-doped La$_{2-x}$Sr$_{x}$CuO$_{4}$ (LSCO) system, the long-range
AFM ordering is completely destroyed at $x\sim 0.02$ and the
superconductivity appears above $x\sim 0.05$ \cite{RMP-70-897}. It is
reported that the long-range AFM ordering in electron-doped Nd$_{2-x}$Ce$%
_{x} $CuO$_{4}$ (NCCO) can extend up to $x\sim 0.13$ \cite{Nat-445-186} or
0.15 \cite{Nat-423-522}. In electron-doped Pr$_{2-x}$Ce$_{x}$CuO$_{4}$
(PCCO) thin films, extensive transport studies strongly suggest a quantum
phase transition at $x\sim 0.16$ \cite{dagan1,dagan2,pcli} and a signature
of static or quasistatic antiferromagnetism up to $x\sim 0.15$ \cite%
{weiqiang}.

For NCCO and PCCO, the optimal doping is around 0.15, while for
electron-doped La$_{2-x}$Ce$_{x}$CuO$_{4}$\ (LCCO), the optimally doped
region shifts to $x\sim 0.09-0.11$ \cite{prb-77-060505,sawa,Wu}. This
difference has been suggested to originate from a smaller antiferromagnetic
exchange interaction in LCCO than that in NCCO and PCCO \cite{sawa}.
However, the magnetic nature of LCCO has never been determined
experimentally, because the $T^{^{\prime }}$-phase LCCO has only been
synthesized in thin film form, where neutron scattering techniques cannot be
used. In this paper, we show that it is possible to determine the magnetic
order in LCCO by using transport measurements.

In films, it has been shown that in-plane angular magnetoresistance (AMR)
measurements can shed light on the magnetic order by probing the spin-charge
coupling. \cite%
{weiqiang,prl-92-227003,prl-83-2813,prb-68-094506,jetp-81-486,YPBCO,jpcm-20-275226,arXiv}%
. In lightly electron-doped Pr$_{1.3-x}$La$_{0.7}$Ce$_{x}$CuO$_{4}$ (PLCCO, $%
x=0.01$) crystals, a fourfold in-plane AMR has been observed, due to a
magnetic-field-induced transition from a noncolinear to colinear Cu-spin
arrangement in adjacent CuO$_{2}$ planes, with the \textquotedblleft
spin-flop\textquotedblright\ easy-axis along the Cu-Cu ([110]) direction and
hard-axis along the Cu-O-Cu ([100]) direction \cite{prl-92-227003}. The
fourfold AMR has also been reported in underdoped NCCO \cite%
{jpcm-20-275226,jetp-81-486} and in PCCO up to $x\sim 0.15$ \cite{weiqiang}.
For PCCO, the temperature at which the fourfold AMR disappears is consistent
with the static or quasistatic AFM ordering temperature determined by
neutron scattering on large crystals \cite{weiqiang}.

In other systems such as underdoped LSCO \cite{prb-68-094506}, YBa$_{2}$Cu$%
_{3}$O$_{6+x}$ (YBCO) \cite{prl-83-2813} and the newly discovered BaFe$%
_{2-x} $Co$_{x}$As$_{2}$ (Ba122) \cite{arXiv}, a twofold in-plane AMR has
been observed. Additionally, the coexistence of twofold and fourfold AMR has
been found in Y$_{0.2}$Pr$_{0.8}$Ba$_{2}$Cu$_{3}$O$_{7-\delta }$ \cite{YPBCO}%
. The twofold behavior in these systems has also been suggested to be
associated with spin ordering. For Ba122, the temperature at which the
twofold AMR disappears coincides with the spin-density-wave (SDW) ordering
temperature \cite{arXiv}.

In this paper, we report the in-plane AMR of LCCO thin films with $%
x=0.06-0.15$. The in-plane AMR in LCCO shows a twofold symmetry, which is
distinct from other electron-doped cuprates where a fourfold symmetry is
found. This suggests that the spin-flop transition does not occur in LCCO.
This twofold AMR disappears at a certain temperature, $T_{D}$. The $T_{D}(x)$
is well above $T_{c}(x)$ for $x=0.06$ ($\sim 110$ K), and decreases with
increasing doping, until it is no longer observed above $T_{c}(x)$ at $%
x=0.15 $. This characteristic $T_{D}(x)$ is suggested to originate from
static AFM ordering or a SDW transition.

%
\begin{figure}[tbp]
\begin{center}
\includegraphics*[bb=41 420 557 749, width=8.5cm,clip]{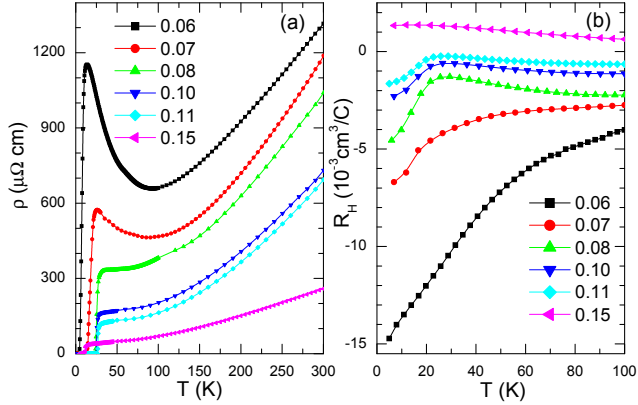}
\end{center}
\caption{(color online) The temperature dependence of resistivity in zero
magnetic field (a) and Hall coefficient (b) in 14 T of LCCO\ thin films with
$x=0.06-0.15$. }
\label{RT}
\end{figure}
%

The $c$-axis oriented LCCO films were deposited directly on (100) SrTiO$_{3}$
substrates by a pulsed laser deposition (PLD) technique utilizing a KrF
excimer laser as the exciting light source. The films were deposited in an
oxygen pressure of $\sim $230 mTorr at $700-750^{\circ }$C. After the
deposition, the films were annealed in vacuum between $10^{-5}-10^{-6}$ Torr
for $15-30$ min to achieve the highest $T_{c0}$ (zero resistance
superconducting transition temperature) and sharpest transition width for
each doping. The samples used for this study are $\sim 2000$ \AA\ and
patterned into a Hall-bar shape with the bridge typically along the $a$
axis. The measurements were carried out using a Quantum Design PPMS 14 T
magnet, and the AMR was measured at temperatures above $T_{c}$ due to the
large in-plane upper critical field.

As shown in Fig. \ref{RT}(a), the resistivity of LCCO films decreases with
increasing $x$ from 0.06 to 0.15. All the films are superconducting at low
temperatures with the optimal doping around $x=0.10-0.11$. Compared with the
LCCO films prepared by a dc magnetron sputtering (MS) method \cite{Wu}, a
PLD technique employing BaTiO$_{3}$ as buffer layer (PLD+BUFFER) \cite{sawa}%
, and a molecular beam epitaxy (MBE) technique \cite{Naito}, our samples
show a broader superconducting region and a lower resistivity in the
underdoped region. The $T_{c0}$ of the optimal doping ($\sim 25$ K)\ in our
films is comparable to that found by the PLD+BUFFER method but slightly
lower than that found by MS and MBE methods. Although different techniques
and preparation processes result in these slight differences, the optimal
doping region and the temperature dependence of resistivity do not change.
To verify the doping concentrations, we also measured the Hall coefficient,
obtained by subtracting the transverse Hall voltage in -14 T from that in 14
T ($\mathbf{H}$ $\perp ab$ plane). The $R_{H}$ gradually changes from
negative to positive with increasing Ce concentration as seen in Fig. \ref%
{RT}(b), consistent with previously reported behavior \cite{sawa,jk2008b}.

%
\begin{figure}[tbp]
\begin{center}
\includegraphics*[bb=56 377 503 745, width=8.5cm,clip]{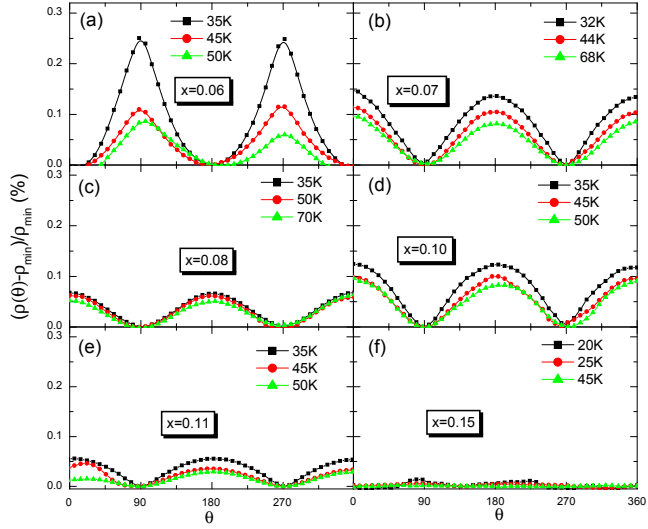}
\end{center}
\caption{(color online) The in-plane angular magnetoresistivity of LCCO
films with $x=0.06-0.15$ at 14 T. $\protect\theta $ is the angle between $%
\mathbf{H}$ and the direction normal to the current ($\protect\theta %
=0^{\circ }$ corresponds to $\mathbf{H\perp I}$ ). Note that here $\mathbf{I}
$ is along the $a$ axis. }
\label{AMR}
\end{figure}
%

For the in-plane AMR measurements, the film was rotated around the $c$ axis
with $\mathbf{H}$ $\parallel ab$. The configurations $\mathbf{H}$ $\parallel
\mathbf{I}$ and $\mathbf{H}$\textbf{\ }$\perp \mathbf{I}$ are referred to $%
\theta =90^{\circ }$ and $0^{\circ }$, respectively. Here $\theta $ is the
angle between $\mathbf{H}$ and the direction normal to $\mathbf{I}$ [see
inset of Fig. \ref{two}(a)]. We define the AMR as $(\rho (\theta )-\rho
_{\min })/\rho _{\min }$ and plot it as a function of $\theta $ for the
films with $x=0.06-0.15$ at 14 T in Fig. \ref{AMR}, where the $\rho _{\min }$
represents the minimum resistivity when $\theta $ changes from $0^{\circ }$
to $360^{\circ }$. Three main features are clearly seen: ($i$) only twofold
symmetry exists in the underdoped and optimal doped LCCO films [Fig. \ref%
{AMR}(a)$-$(e)], and it is not observed above $T_{c}$\ at $x=0.15$ [Fig. \ref%
{AMR}(f)];($ii$) the peak of AMR for $x=0.06$ appears at $\theta =90^{\circ
} $ and $270^{\circ }$ corresponding to $\mathbf{H\parallel I}$ at certain
temperatures (discussed below), while it shifts $90^{\circ }$ for other
doping levels, appearing when $\mathbf{H\perp I}$; ($iii$) the magnitude of
the anisotropic in-plane AMR, $(\rho _{\max }-\rho _{\min })/\rho _{\min }$,
is $\sim $ $0.01\%-0.1\%,$ and decreases with increasing temperature. In
PCCO, the anisotropic AMR is of the same magnitude as in LCCO, but shows
fourfold symmetry due to the anisotropic (fourfold) spin-flop field \cite%
{weiqiang}.

%
\begin{figure}[tbp]
\begin{center}
\includegraphics*[bb=46 269 552 753, width=8.5cm,clip]{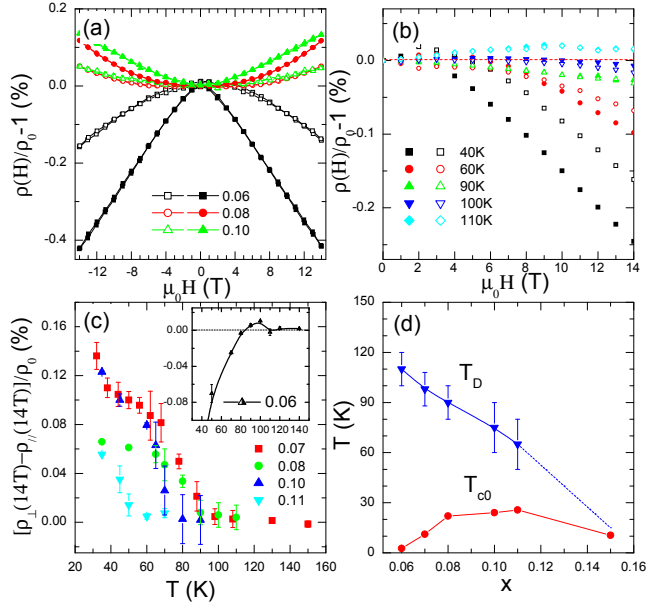}
\end{center}
\caption{(color online) (a) The field dependence of the in-plane
magnetoresistivity [open symbols LMR ($\mathbf{H\parallel I}$), solid
symbols TMR ($\mathbf{H\perp I}$)] of LCCO with $x=0.06,0.08$ and 0.10 at 35
K. (b) the LMR and TMR of $x=0.06$ at different temperatures. (c) The
difference between LMR and TMR at 14 T. (d) $T_{D}(x)$ and $T_{c0}(x)$ of
the LCCO. $T_{D}$ and $T_{c0}$ represent the temperature where the twofold
AMR disappears and the zero-resistance superconducting transition
temperature, respectively. }
\label{RH}
\end{figure}
%

It should be noted that the data shown in Fig. \ref{AMR} are measured at 14
T, but the twofold AMR is found with smaller magnitude at lower magnetic
fields. In Fig. \ref{RH}(a), magnetoresistivity curves at $\theta =90^{\circ
}$\ ($\mathbf{H}$ $\parallel \mathbf{I}$, LMR) and $0^{\circ }$\ ($\mathbf{H}
$\textbf{\ }$\perp \mathbf{I}$, TMR) at 35 K for three dopings are plotted
as a function of the magnetic field. The difference between TMR and LMR
increases with increasing $H$ so that the maximal AMR signal appears at 14
T. The twofold behavior is clear at lower temperatures, whereas at higher
temperatures, it is not distinguishable due to smaller signal and much
higher temperature fluctuations in the Quantum Design rotator. To probe the
temperature at which the twofold behavior disappears ($T_{D}$), we used a
Quantum Design resistivity sample stage and measured the magnetoresistivity
of the samples with $x=0.06-0.11$ under the fixed configurations of $\mathbf{%
H\perp I}$ and $\mathbf{H\parallel I}$.

In Fig. \ref{RH}(b), we show the field dependence of the TMR (solid symbols)
and the LMR (open symbols) for $x=0.06$ sample at different temperatures. At
low temperatures, the TMR is more negative than the LMR. With increasing
temperature, the TMR becomes larger than the LMR, and finally they overlap
with each other. The difference between TMR and LMR at 14 T, $\delta \rho
(14T)=[\rho _{\perp }(14T)-\rho _{//}(14T)]/\rho _{0}$ ($\rho _{0}$ is the
zero field resistivity), is shown in the inset of Fig. \ref{RH} (c). At $%
\sim $ 80 K, the $\delta \rho (14T)$ changes from negative to positive.
Thus, the peak of the AMR of $x=0.06$ sample at $T>80$ K shifts $90^{\circ }$%
, showing a similar shape as that for higher doping levels. The $\delta \rho
(14T)$ almost reaches zero above $T_{D}\sim $110 K, i.e., the twofold AMR
disappears above 110 K. Using the same method, we also obtained $T_{D}$ for
other LCCO films. As shown in Fig. \ref{RH} (c), the samples with $%
x=0.07,0.08,0.10$ and 0.11 show the same behavior: the difference between
TMR and LMR gradually decreases with increasing temperature and almost
disappears above $T_{D}(x)$. The $T_{D}(x)$ decreases with increasing
doping, until it is not observed above $T_{c0}(x)$ at $x=0.15$ as seen in
Fig. \ref{RH}(d). \textit{This is the most important finding in this study.}

We shall now discuss the possible origin of this $T_{D}(x)$. In Ba122, the
twofold AMR has been ascribe to SDW ordering \cite{arXiv}. We note that
though a fourfold AMR has been reported in PCCO, it is likely that a twofold
AMR also exists, as seen in the Figure 4 of Ref.\cite{weiqiang}. Moreover,
the twofold and fourfold AMR seems to disappear roughly at the same
temperature, so the twofold AMR in PCCO may also be caused by the static or
quasi-static AFM ordering \cite{weiqiang}. Therefore, it is most likely that
the twofold AMR in LCCO originates from an AFM or SDW order.

Other explanations based on magnetic ordering should also be considered.
Ando \textit{et al. }\cite{prl-83-2813} suggested that field-induced
ordering of stripes in YBCO resulted in the anisotropy of the in-plane
magnetoresistivity. However, there is no clear evidence for stripes in
electron-doped cuprates. Another possibility is based on the orthorhombic
distortion. In the AFM ordered state, the crystal structure of lightly doped
YBCO has a small orthorhombic distortion, which may lead to the in-plane
anisotropic magnetoresistance in a magnetic field \cite{prl-85-474}. In
cuprates containing only CuO$_{2}$\ planes, a perfect tetragonal structure
forbids this anisotropy \cite{prl-78-535}. For LSCO, there is a
tetragonal-to-orthorhombic transition and the structure is orthorhombic in
the AFM ordered state \cite{RMP-70-897}. The electron-doped cuprates are
known to be in tetragonal structure. If the apical oxygens are not fully
removed after annealing \cite{prb-70-064513}, a local orthorhombic
distortion may occur in LCCO as in LSCO, resulting in a twofold AMR. The
LCCO films containing apical oxygens should be in $T$-phase structure \cite%
{Naito}. The magnitude of AMR in underdoped LCCO is comparable to that in
underdoped YBCO and LSCO. If the twofold AMR is caused by the orthorhombic
distortion, the $T$-phase peaks should be detectable by X-ray diffraction,
However,\ our X-ray diffraction data only show strong $T^{^{\prime }}$-phase
(00\textit{l}) peaks.

%
\begin{figure}[tbp]
\begin{center}
\includegraphics*[bb=0 0 460 284, width=8.5cm,clip]{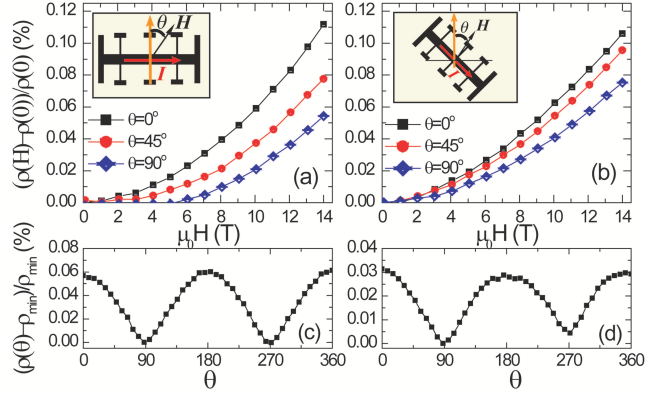}
\end{center}
\caption{(color online) Magnetoresistivity measured at $\protect\theta %
=0^{\circ },45^{\circ },$ and 90$^{\circ }$ with $\mathbf{I}$ along the $a$
axis (a) and along the diagonal direction (b) for the samples with $x=0.08$
at 35 K. (c) and (d) are the corresponding AMR for these two configurations.}
\label{two}
\end{figure}
%

We should also discuss possible explanations without magnetic ordering.
Firstly, for our thin films, the in-plane anisotropy is of magnitude $\sim $%
0.1\%, one may expect that if there is a small angle between the field and
the $ab$ plane, the \textit{c}-axis field component can cause the
difference. However, no asymmetric component can be resolved from the field
sweep as shown in Fig. \ref{RH} (a), so we can rule out this possibility.

Secondly, it is possible that the twofold AMR is caused by an extrinsic
difference of resistivity for $\mathbf{H\parallel I}$ vs $\mathbf{H\perp I}$%
. However, if this were true, we should see twofold AMR in the films with $%
x=0.15$ as well, and we do not. Moreover, we patterned two bridges, one
along the $a$ axis and another $45^{\circ }$ offset, on the same sample and
measured their in-plane magnetoresistivity simultaneously. As seen in Fig. %
\ref{two}, the $\theta $ is defined as the angle between $\mathbf{H}$ and
the $a$ axis. It is clear at 35 K that for both configurations, the
magnetoresistivity of the $x=0.08$ sample at $\theta =45^{\circ }$ is
located between that at$\ \theta =0^{\circ }$ and $90^{\circ }$ as seen in
Fig. \ref{two}(a) and (b). From the corresponding AMR [Fig. \ref{two}(c) and
(d)], we can see that for the second configuration (bridge along [110]), the
AMR at $\theta =45^{\circ }$ ($\mathbf{H\perp I}$) is almost equal to that
at $\theta =135^{\circ }$ ($\mathbf{H\parallel I}$). We find the same result
for other dopings and at other temperatures below $T_{D}$. So the origin of
the twofold AMR is not due to the angle between $\mathbf{H}$ and $\mathbf{I}$%
.

Thirdly, recent Nernst experiments have revealed that a large Nernst signal
can exist at temperatures above $T_{c}$ in both electron- and hole-doped
cuprates, suggesting an extended phase fluctuation region \cite%
{prb-73-024510,prb-76-174512}. However, this fluctuation region is
dome-shaped, and does not monotonously decrease as doping increases.
Moreover, in electron-doped PCCO and NCCO, the phase fluctuation temperature
extends only up to $\sim 30$ K \cite{prb-76-174512,prb-73-024510}. Thus, the
$T_{D}(x)$ does not originate from superconducting phase fluctuations.
Tunneling experiments in PCCO observed that a normal-state gap vanished at a
certain temperature $T^{\ast }$ \cite{prl-94-187003}. This characteristic $%
T^{\ast }$ is greater than $T_{c}$ for the underdoped region and follows $%
T_{c}$ on the overdoped side. However, the $T^{\ast }$ is lower than 30 K,
so it is not responsible for the twofold AMR.

Fourthly, there are two kinds of charge carriers in LCCO \cite{jk2007}.
However, even if this could result in the in-plane anisotropy, it has been
found that the magnetoresistance caused by the two-band feature is strongest
near the optimal doping \cite{jk2008b}, similar to PCCO \cite{prb-73-024510}%
. So two kinds of charge carriers are unlikely to be the origin of the
twofold AMR.

Thus, we suggest that the $T_{D}(x)$ is originates from a static or
quasi-static AFM or SDW ordering. We note that the $R_{H}(T)$ of underdoped
and optimally doped LCCO films shows a downturn at certain doping dependent
temperatures, but these temperatures are much lower than the $T_{D}(x)$. In
PCCO films, the $R_{H}(T)$ also shows a downturn in underdoped and optimally
doped regions at certain tempertures\cite{dagan1}, and these temperatures
are also lower than the AFM temperatures \cite{weiqiang}. The origin of the
downturn of $R_{H}(T)$ is not understood at this time, but it must be
something other than the AMF ordering. In the LCCO films, the disappearance
of $T_{D}(x)$, the insulator-to-metal transition \cite{jk2008a}, and the
formation of a large holelike Fermi surface \cite{jk2008b} all occur at $%
x\sim 0.15,$\ suggesting a relation among these behaviors. According to the
above discussion, we infer that the disappearance of an AFM or SDW ordering
causes all these behaviors.

In summary, we investigated the in-plane angular magnetoresistance (AMR) of
La$_{2-x}$Ce$_{x}$CuO$_{4}$\ (LCCO) thin films ($x=0.06-0.15$), and observed
a twofold symmetry. Unlike other electron-doped cuprates, a fourfold AMR
caused by a spin-flop transition is not observed in the LCCO system. The
twofold AMR disappears above a certain temperature, $T_{D}$. The $T_{D}(x)$
decreases with increasing doping ($T_{D}\sim 110$ K for $x=0.06$), falling
below $T_{c0}(x)$ at $x=0.15$. This newly disclosed characteristic $T_{D}(x)$
is suggested to originate from an static AFM or SDW ordering.

The authors would like to thank N. Butch, S. Saha, and J. Paglione for
fruitful discussions. K. J. acknowledges W. Yu for technical help and useful
discussions. This work is supported by the NSF under DMR-0653535.


\end{document}